\documentstyle[12pt,dina4p,graphicx,amssymb]{article}
\begin{document}

\vspace{1cm}

\newcommand{\ra}{\rightarrow}
\newcommand{\as}{\mbox{$\alpha_{\displaystyle  s}$}}
\def\pom{{I\!\!P}}
\def\reg{{\cal R}}
\def\sm0{{ o}}
\def\pt{{\rm P_t}}
\def\pbarp{{\rm \overline{p}p}}
\def\mev{{\rm \ MeV}}
\def\gev{{\rm \ GeV}}
\def\gpom{\gamma^*\!-\!\pom}

\begin{titlepage}
\title{
{\bf Event Shape Analysis \\
of Deep Inelastic Scattering Events \\
with a Large Rapidity Gap at HERA}
\author{ZEUS Collaboration}
}
\date{}
\maketitle

\vspace{5cm}
\begin{abstract}
A global event shape analysis of the multihadronic
final states observed in neutral current deep inelastic scattering events 
with a large rapidity gap with respect to the proton direction 
is presented. The analysis
is performed in the range $5 \leq Q^2 \leq 185\gev^2$ and
$160 \leq W \leq 250\gev$, where $Q^2$ is the virtuality 
of the photon and $W$ is the virtual-photon proton centre
 of mass energy. Particular
emphasis is placed on the dependence of the shape variables, measured in the
$\gamma^*\!-$pomeron rest frame, on the mass of the hadronic final state, $M_X$.
With increasing $M_X$ the multihadronic final state becomes more collimated 
and planar. The experimental results are compared with several models which 
attempt to describe diffractive events. The
broadening effects exhibited by the data require in these models a significant
gluon component of the pomeron. 

\end{abstract}

\vspace{-19cm}
{\noindent
 DESY 97-202 \newline
 October 1997}

\setcounter{page}{0}
\thispagestyle{empty}   
\pagenumbering{Roman}
\def\3{\ss}
\parindent0.cm
\parskip 3mm plus 2mm minus 2mm

\newpage

\begin{center}                                                                                     
{                      \Large  The ZEUS Collaboration              }                               
\end{center}                                                                                       
  J.~Breitweg,                                                                                     
  M.~Derrick,                                                                                      
  D.~Krakauer,                                                                                     
  S.~Magill,                                                                                       
  D.~Mikunas,                                                                                      
  B.~Musgrave,                                                                                     
  J.~Repond,                                                                                       
  R.~Stanek,                                                                                       
  R.L.~Talaga,                                                                                     
  R.~Yoshida,                                                                                      
  H.~Zhang  \\                                                                                     
 {\it Argonne National Laboratory, Argonne, IL, USA}~$^{p}$
% ...
\newpage                                                                                           
% ...
$^{  23}$ now at ATLAS Collaboration, Univ. of Munich \\                                           
% ...
$^{  32}$ partially supported by the Foundation for German-Russian Collaboration                   
DFG-RFBR \\ \hspace*{3.5mm} (grant no. 436 RUS 113/248/3 and no. 436 RUS 113/248/2)\\              
% ...
                                                           %
                                                           %
\newpage   
                                                           %
                                                           %
\begin{tabular}[h]{rp{14cm}}                                                                       
$^{a}$ &  supported by the Natural Sciences and Engineering Research                               
          Council of Canada (NSERC)  \\                                                            
$^{b}$ &  supported by the FCAR of Qu\'ebec, Canada  \\                                            
$^{c}$ &  supported by the German Federal Ministry for Education and                               
          Science, Research and Technology (BMBF), under contract                                  
          numbers 057BN19P, 057FR19P, 057HH19P, 057HH29P, 057SI75I \\                              
$^{d}$ &  supported by the MINERVA Gesellschaft f\"ur Forschung GmbH,                              
          the German Israeli Foundation, and the U.S.-Israel Binational                            
          Science Foundation \\                                                                    
$^{e}$ &  supported by the German Israeli Foundation, and                                          
          by the Israel Science Foundation                                                         
  \\                                                                                               
$^{f}$ &  supported by the Italian National Institute for Nuclear Physics                          
          (INFN) \\                                                                                
$^{g}$ &  supported by the Japanese Ministry of Education, Science and                             
          Culture (the Monbusho) and its grants for Scientific Research \\                         
$^{h}$ &  supported by the Korean Ministry of Education and Korea Science                          
          and Engineering Foundation  \\                                                           
$^{i}$ &  supported by the Netherlands Foundation for Research on                                  
          Matter (FOM) \\                                                                          
$^{j}$ &  supported by the Polish State Committee for Scientific                                   
          Research, grant No.~115/E-343/SPUB/P03/002/97, 2P03B10512,                               
          2P03B10612, 2P03B14212, 2P03B10412 \\                                                    
$^{k}$ &  supported by the Polish State Committee for Scientific                                   
          Research (grant No. 2P03B08308) and Foundation for                                       
          Polish-German Collaboration  \\                                                          
$^{l}$ &  partially supported by the German Federal Ministry for                                   
          Education and Science, Research and Technology (BMBF)  \\                                
$^{m}$ &  supported by the Fund for Fundamental Research of Russian Ministry                       
          for Science and Edu\-cation and by the German Federal Ministry for                       
          Education and Science, Research and Technology (BMBF) \\                                 
$^{n}$ &  supported by the Spanish Ministry of Education                                           
          and Science through funds provided by CICYT \\                                           
$^{o}$ &  supported by the Particle Physics and                                                    
          Astronomy Research Council \\                                                            
$^{p}$ &  supported by the US Department of Energy \\                                              
$^{q}$ &  supported by the US National Science Foundation \\                                       
\end{tabular}                                                                                      
                                                           %
                                                           %
\end{titlepage}
 
\newpage
\parskip 3mm plus 2mm minus 2mm
\parindent 5mm
\pagenumbering{arabic}
\setcounter{page}{1}
\normalsize

\section{Introduction}
At HERA a class of neutral current (NC) deep inelastic scattering (DIS) 
events has been observed
which is characterized by a large rapidity gap (LRG) between the proton beam 
direction and a multi-particle final state \cite{zlrg0,h11}. The $Q^2$ dependence of the LRG event fraction points to a leading
twist production mechanism. The rate of LRG events is most commonly accounted for by introducing ``diffractive'' processes, which proceed via the 
t-channel exchange of a colour singlet object with
quantum numbers of the vacuum, called the pomeron ($\pom$).
However, the nature of
the pomeron is at present far from clear. Ingelman and Schlein \cite{gips}
assumed that the pomeron emitted from the proton behaves like a hadron
and suggested that it could have a partonic
substructure which could be probed by a hard
scattering process. The UA8 experiment at CERN later observed events containing
two high-$p_T$ jets in $\overline{p}p$ interactions which were tagged with
leading protons (or antiprotons).
This observation was explained in terms of a partonic structure
in the pomeron \cite{ua8}.\\
Jet production in LRG events at HERA in the photoproduction regime \cite{zlrg3} 
and the pattern of scaling violations in diffractive DIS \cite{h1diff97} led
to the conclusion that in order to interpret the data within the
Ingelman-Schlein approach, the pomeron must have
a high gluon content.\\
Various models for diffractive scattering can be used to describe the 
hadronic final state in LRG events \cite{models}. 
The theoretical predictions range from a calculation of gluon
pair exchange \cite{dl,nz}, to a phenomenological description of pomeron
quark and gluon densities \cite{pompyt,rapgap} and to a boson-gluon
fusion scheme, where the resulting quark pair evolves into a 
colour singlet state \cite{sci,sci2}.
Recently, calculations for two- and three-jet
production in DIS rapidity gap events have been presented, 
in which a scalar pomeron is considered
to have a pointlike coupling to quark and gluon pairs \cite{vermaseren}.\\ 
In order to obtain a deeper understanding of multihadronic final states
with a large rapidity gap, we study shape variables
in the centre of mass frame of the observed hadronic system (of mass $M_X$),
interpreting this as the $\gpom$ c.m. system, and investigate their dependence
upon $M_X$. This
 is in analogy with the  studies of
shape variables in $e^+ e^-$ annihilation as a function of $\sqrt{s}$
\cite{barreiro} and to those
which led to the interpretation of three-jet events as a
consequence of gluon bremsstrahlung \cite{tasso,pluto,markj,jade}. It
has also been applied to the dependence of fixed target antineutrino data as a  
function of $W$ \cite{fernandez}.

\section{Experimental setup}

The data presented here, taken with the ZEUS detector at HERA in 1994,
correspond to a luminosity of 2.57$\pm$0.04~~${\rm pb^{-1}}$.
HERA operated with 153 colliding bunches of 820~GeV protons and 27.5 GeV
positrons.
Additional unpaired positron and proton bunches
circulated,
which were used to determine beam related backgrounds.
 
A description of the ZEUS detector can be
found in~\cite{sigtot,Detector}.
The components used in this analysis are briefly discussed here.
The uranium-scintillator calorimeter (CAL)~\cite{CAL} covers
99.7\% 
of the total solid angle. 
It consists of the forward calorimeter (FCAL) covering the range
in pseudorapidity\footnote{The ZEUS coordinate system
is defined as right-handed with the $Z$ axis pointing in the proton beam
direction, and the $X$ axis horizontal, pointing towards the centre of
HERA. The pseudorapidity is defined as
$\eta=-\ln(\tan\frac{\theta}{2})$,
where $\theta$ is the polar angle with respect to the proton di\-rec\-tion.}
$4.3<\eta<1.1$, the barrel calorimeter (BCAL) covering $1.1<\eta<-0.75$
and the rear calorimeter (RCAL) covering $-0.75<\eta<-3.8$.
Each calorimeter part is segmented
into electromagnetic (EMC) and hadronic (HAC) sections.
Each section is further subdivided into cells of typically
$5 \times 20$ cm$^2$ ($10 \times 20$~cm$^2$
in the RCAL) for the EMC
and $20 \times 20$~cm$^2$ for the HAC sections.
Under test beam conditions the calorimeter has an energy resolution of
$\sigma/E$~=~18\%/$\sqrt{E ({\rm GeV})}$
for electrons and
$\sigma/E$~=~35\%/$\sqrt {E({\rm GeV})}$ for hadrons.
The timing resolution of a calorimeter cell is less  than 1 ns for energy
deposits greater than $4.5\gev$.
In order to minimize the effects of noise due to the uranium radioactivity
 on the energy measurements
all EMC(HAC) cells with an energy deposit of less than 60(110)~MeV
are discarded from the analysis. 
 
The tracking system consists of a vertex detector (VXD)~\cite{VXD},
a central tracking chamber (CTD)~\cite{CTD}, and a rear tracking
detector (RTD)~\cite{ftd,Detector}
enclosed in a 1.43 T solenoidal magnetic field. The interaction vertex
is measured with a typical resolution along and transverse to the beam direction
of $0.4$~cm and $0.1$~cm respectively. 
 
The position of positrons scattered at small angles
with respect to the positron beam direction
is measured using the small angle rear tracking detector (SRTD)
 which is attached to the front face of the RCAL.
The SRTD \cite{SRTD} consists of two planes of scintillator strips, each one 
1 cm wide and 0.5 cm thick, arranged in orthogonal directions and read out
via optical fibers and photo-multiplier tubes. 
 
The luminosity is measured via the Bethe-Heitler process,
 $ep \rightarrow e \gamma p$,
using a lead-scintillator calorimeter (LUMI)~\cite{LUMI}
which accepts photons at angles~$\le$~0.5~mrad with respect to the
positron beam direction and is located at $Z=-107$~m.
 
\section{Reconstruction and kinematic variables}
 
The kinematic variables used to describe deep inelastic $ep$ scattering
\[
     e~(k)~+~ p~(P) \rightarrow e~(k')~+~X
\]
are the following:

\noindent the negative squared four-momentum 
transfer carried by the virtual photon
\[
      Q^2=-q^2=-(k~-~k')^2,
\]
the Bjorken $x$ variable
\[
     x =\frac{Q^2}{2P\cdot q},
\]
the fractional energy transfer to the hadronic final state
\[
     y =\frac{P\cdot q}{P\cdot k},
\]
and the square of the centre-of-mass energy of the virtual-photon proton system 
($\gamma^*p$)
\[
   W^2=(q+P)^2=\frac{Q^2(1-x)}{x}+M_p^2,
\]
where $M_p$ is the proton mass.

These variables, only two of which are independent, can be determined
either from the scattered positron or from the
hadronic system. They can also be determined from a mixed set of variables,
the so-called double angle method \cite{bent}. In this analysis,
the scattered positron method is used because it provides a better resolution in
the kinematic range under study. The variables $Q^2$ and $y$, 
calculated from 
the scattered positron variables, are given by the expressions
\[
Q_e^2 = 2E_eE^{\prime}_e(1+cos \theta^{\prime}_e),
\]
and
\[
y_e = 1 - \frac{E^{\prime}_e}{2E_e}(1-cos \theta^{\prime}_e),
\]
where $E_e$ is the incident positron energy and $E^{\prime}_e$ and
 $\theta^{\prime}_e$ denote the energy and angle, with respect to the 
proton direction, of the scattered positron. 

To identify the neutral current
DIS events the quantity $\delta$ is also used:
\[
 \delta=\sum_i (E_i - p_{Z\,i}),
\]
where $E_i$ denotes the energy assigned to the signal in calorimeter cell $i$ and
$p_{Z\,i}=E_i\cos{\theta_i}$, $\theta_i$ being the polar angle of the cell.
The sum is running over all calorimeter cells. For fully contained
DIS events, and neglecting detector smearing
effects, $\delta=2\cdot E_e$.

In order to describe the kinematics of diffractive scattering
\[
e~(k) + p~(P) \rightarrow e~(k') + p~(P') + X
\]
the following two additional variables are introduced: the squared four-momentum
transfer at the proton vertex,
\[
t = (P-P^{\prime})^2,
\]
whose absolute magnitude is small compared to $Q^2+M_X^2$
in the diffractive processes selected in this analysis, 
and $x_{\pom}$, the fraction
of the proton momentum carried by the pomeron, defined as
\[
 x_{\pom} = \frac{(P-P')\cdot q}{P\cdot q}=\frac{M_X^2+Q^2-t}{W^2+Q^2-M_p^2}
 \simeq  \frac{M_X^2+Q^2}{W^2+Q^2},
\]
where $M_X$ is the mass of the hadronic system. 

In addition, the variable $\beta$ is defined as
\[
\beta=\frac{Q^2}{2(P-P')\cdot q} = \frac{x}{x_{\pom}} =
\frac{Q^2}{M_X^2+Q^2-t} \simeq \frac{Q^2}{M_X^2+Q^2},
\]
which can be interpreted as the fraction of the pomeron momentum carried by the
struck quark.

\section{Trigger and data selection criteria}

Events were filtered online by a three level trigger system~\cite{Detector}.
At the first level \cite{CALFLT}, DIS events are selected by requiring a
minimum energy deposition in the EMC section of the CAL. The threshold varies
between 3.4 GeV (in the RCAL) and 4.8 GeV (in the BCAL). At the second level (SLT),
non-$ep$ background is further reduced by using the measured times of energy
deposits and the summed energies from the calorimeter. 
The events are accepted if\\
  $\delta_{SLT} \equiv \displaystyle \sum_i E_i(1-\cos\theta_i) > (24 - 2E_{\gamma})\gev$,
where $E_i$ and $\theta_i$ are the energies and polar angles 
of calorimeter cells, and $E_{\gamma}$
is the energy deposit measured in the LUMI photon calorimeter.
This requirement on $\delta_{SLT}$ does not reject events with initial 
state QED radiation.

The full event information is available at the third level trigger (TLT).
Tighter timing cuts as well as algorithms to remove beam halo and cosmic
muons are applied. 
The quantity $\delta_{TLT}$ was determined in the same way as
$\delta_{SLT}$. The events were required to have
\mbox{$\delta_{TLT}>(25 - 2E_{\gamma})\;$GeV}.
Finally, events were accepted if
a scattered positron candidate of energy greater than 4~GeV was found.
In total 900,853 NC DIS candidates satisfied the above trigger conditions.
 
DIS events were selected off-line by looking for scattered
positron candidates recognized using the pattern of energy deposition in the
calorimeter. The efficiency of the positron identification algorithm, 
as well as the procedures to incorporate information from the SRTD 
in order to improve the
positron energy and scattering angle measurements are thoroughly
discussed in our recent publication on the measurement of the 
proton structure function $F_2$ \cite{f2}.

The off-line cuts used to select multihadronic DIS events are the
following:
\begin{itemize}
\item $E'_e> 8\gev$.
\item $5\gev^2 \leq Q^2_{e} \leq 185\gev^2$.
\item $y_{e}\leq 0.95$. This cut removes fake positrons found in the
FCAL.
\item $35\gev \leq \delta \leq 65\gev$. This cut removes events with
large initial state QED radiation and further reduces the background from
photoproduction.
\item The events are required to have a reconstructed vertex. This cut
suppresses beam-gas background.
\item Events with a scattered positron impact point in the
RCAL inside a box of 26$\,$cm$\times 26\,$cm around the beam pipe are rejected.
This guarantees full containment of the shower in the calorimeter.
\end{itemize}

A total of 294,527 events were selected this way. 

The mass of the final state hadronic system, $M_X$, can be reconstructed from the
energy deposited in the calorimeter cells,
excluding those assigned to the outgoing positron.
This reconstruction is sensitive to calorimeter noise effects. To improve it,
especially at low masses, we removed isolated cells
with energy below 140 (160) MeV for EMC (HAC) cells.
The following formula was used for the mass reconstruction:
\[
   M_X^{rec}\equiv A_1 \cdot \sqrt{(\sum E)^2 - (\sum p_X)^2 - (\sum p_Y)^2 -
(\sum p
_Z)^2}+ A_0.
\]
The coefficients $A_0$ and $A_1$ correct for the effects of energy loss
in the inactive material and energy deposits below the threshold.
Their values, $A_0=1.4$~GeV and $A_1=1.21$, 
were determined from the Monte Carlo
simulation so as to give the best estimate of the true invariant mass
in diffractive events in the kinematic region of the analysis.
In this region, the resolution in $M_X^{rec}$ is approximately $\sigma(M_X^{rec})/M_X^{rec} \approx 60\%/\sqrt{M_X^{rec}}$. 

Diffractive events are characterized by a small value of $M_X$. However a
simple selection based on $M_X^{rec}$ is not enough to select diffractive 
events because the non-diffractive background would be too high even at
relatively low values of the mass. This is illustrated in
Fig.~\ref{fig:feyn-lnmx2}b where the $\ln M_X^{2~rec}$ 
distribution for the DIS sample with and without a forward rapidity gap 
requirement
is presented along with a comparison with non-diffractive Monte Carlo expectations
normalized to the integrated luminosity. Therefore,
to reduce such background and extend the range of mass, a selection based
on the presence of a forward rapidity gap is applied. The following cuts
define the final sample for the analysis:
\begin{itemize}
\item $\eta_{max}^{rec} \leq 1.8$, where $\eta_{max}^{rec}$
is defined as the pseudorapidity
of the most forward condensate with an energy above $400\mev$. 
A condensate is an energy deposit in contiguous cells above $200\mev$ in 
the calorimeter.
In Fig.~\ref{fig:feyn-lnmx2}c the $\eta_{max}^{rec}$
distribution for the full DIS data sample as well
as for that with $7 \leq M_X^{rec} \leq 25~GeV$ (see below) 
is presented along with a comparison
with non-diffractive Monte Carlo expectations normalized to the total integrated
luminosity.
\item $7 \leq M_X^{rec} \leq 25\gev$. The lower limit rejects events
where the phase space for QCD radiation is limited. 
The upper cut is dictated mainly by the lack of statistics at high 
masses, a consequence of the $\eta_{max}^{rec}$ requirement.
\item $W \geq 160\gev$. This cut helps to suppress the non-diffractive DIS
background \cite{kowa}. 
\item more than three condensates. This requirement ensures that an event
plane for the hadronic final state $X$ can be determined.
\end{itemize}

A total of 2748 events are selected by these requirements. 
This sample is estimated
to have a contamination from beam-gas background below $1\%$, 
based on the distribution of events in 
unpaired bunches. The background due to photoproduction is
estimated to be less than $3\%$ 
from Monte Carlo studies. Both backgrounds have
been neglected in the analysis.

The resulting reconstructed values of $x_{\pom}$
are below 0.01, thus the selected sample is expected to be due to pomeron
exchange~\cite{f2d4,miraf}.
The remaining
non-diffractive DIS background varies from $5\%$ at the lower end to
$10\%$ at the upper end of the mass range defined above 
(see Fig.~\ref{fig:feyn-lnmx2}b).
The background calculation
has been checked by reweighting the exponential fall-off in the Monte Carlo 
sample in order to agree with
estimates from fits to the data
for $\ln M_X^{2~rec} \lesssim 7$ with $M_X^{rec}$ in GeV \cite{chema}. The two calculations agree within errors.
This contribution will be subtracted on a bin-by-bin basis in the differential
event shape distributions. 
\newpage

\section{ Monte Carlo simulation }
 
Monte Carlo (MC) event simulation is used to correct for detector
acceptance and smearing effects. The detector simulation is based on
the GEANT program \cite{geant} and incorporates our understanding of
the detector and the trigger performance.
 
Events from standard non-diffractive DIS processes with first 
order electroweak 
corrections were generated with the HERACLES 4.5 programme \cite{herac}.
It was interfaced to
ARIADNE \cite{ari} version 4.08 for modeling the QCD cascade according to
the colour dipole model \cite{codi} that includes the boson
gluon fusion diagram, denoted hereafter by CDMBGF. The fragmentation into
hadrons was performed with the Lund fragmentation scheme as
implemented in JETSET 7.4 \cite{jetse}.

\begin{figure}\vspace{-2cm}
\centerline{\includegraphics[height=25cm]{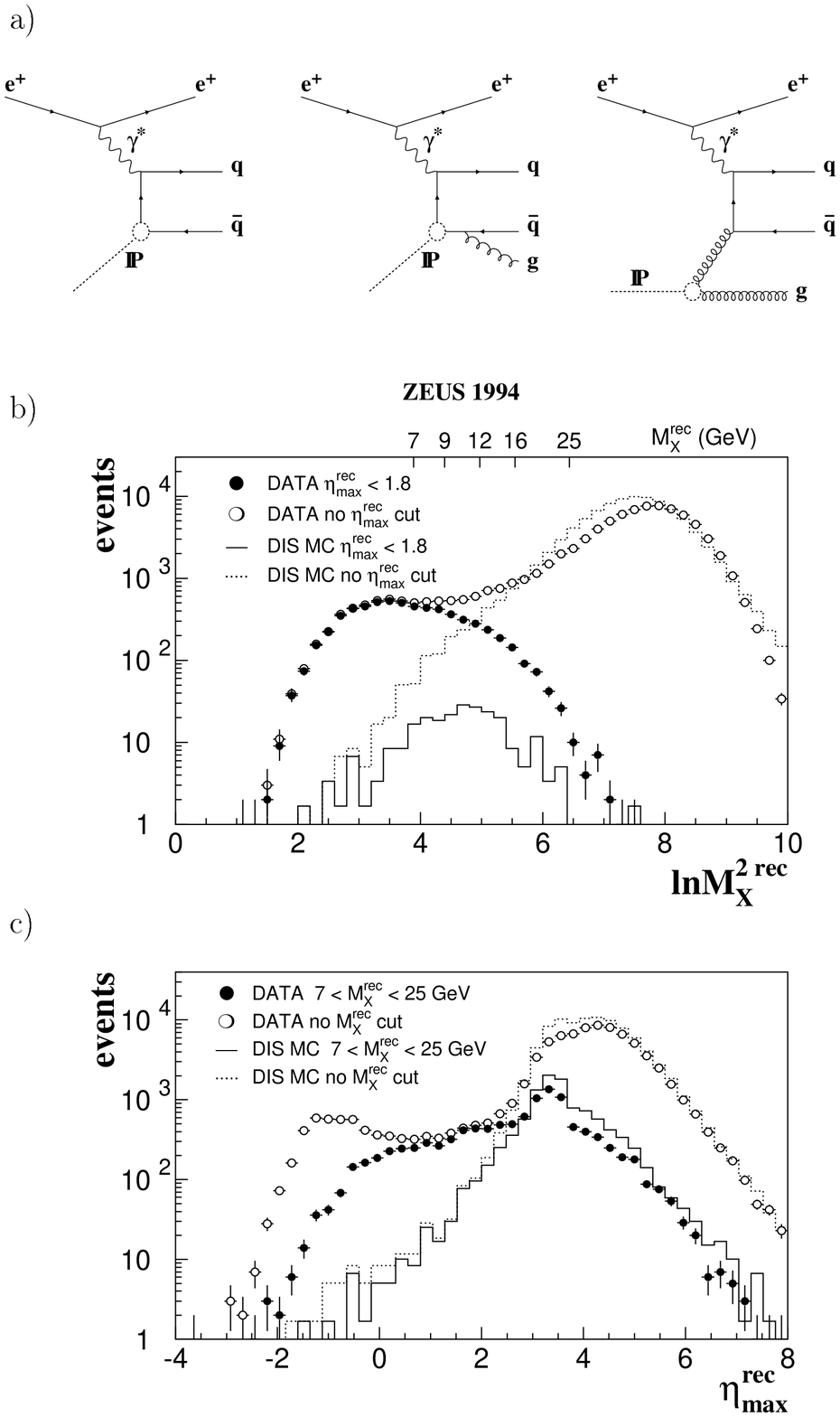}}\vspace{-2cm}
\caption{a) Feynman diagrams for processes in the RAPGAP and VBLY models,
b) lnM$^{2\,rec}_X$ distributions for the complete DIS as well as
large rapidity gap samples, along with MC predictions generated 
within the CDMBGF model
and normalized to the integrated luminosity of the data, c) $\eta_{max}^{rec}$ distributions
for the complete DIS sample as well as for the restricted mass range 
$7 < M_X^{rec}<25\gev$, along with MC predictions as in b).}
\label{fig:feyn-lnmx2}
\end{figure}

In order to model the diffractive hadronic final states
we have considered various models for the pomeron. POMPYT
\cite{pompyt} is a Monte Carlo implementation of factorisable models for
high energy diffractive processes, where, within the framework
provided by PYTHIA \cite{p}, the initial state proton emits a pomeron 
according to the Donnachie-Landshoff flux \cite{dola}. The pomeron
constituents take part in a hard scattering process
with the virtual photon
(see first diagram of Fig.~\ref{fig:feyn-lnmx2}a).
 The quark density in the pomeron is chosen to be of the type 
$\beta f(\beta)\propto \beta(1-\beta)$. 
In POMPYT the  
final states come from the fragmentation of a $q\bar{q}$
pair, and QCD effects (e.g. QCD Compton, second diagram of 
Fig.~\ref{fig:feyn-lnmx2}a)  are approximated by the leading log approximation
for parton showering. 

RAPGAP \cite{rapgap} represents an improvement over POMPYT in that a gluon
density in the pome\-ron is also considered, thus giving rise to final states
produced by a photon gluon mechanism $\gamma^*g \rightarrow q\bar{q}$
(see third diagram in Fig.~\ref{fig:feyn-lnmx2}a).
In the version we have used, the Streng parameterization for the 
pomeron flux \cite{streng}
was used and the quark and gluon densities
in the pomeron satisfy the momentum sum rule. Also, QCD effects are
better modeled within the colour dipole model \cite{codi} as implemented in ARIADNE \cite{ari} than by the leading log parton shower approximation. 
The colour dipole approximation turns out to be exact in the limit of massless quarks for the
process of gluon emission from a $q\bar{q}$ pair.
 First order electroweak corrections
are taken into account with the programme HERACLES \cite{herac}.

A different approach has recently been proposed by VBLY \cite{vermaseren}.
Exact ${\cal O}(\as)$ calculations of two- and three-jet production 
in diffractive DIS are performed. The calculations assume that the 
pomeron is emitted at the proton vertex according to the Donnachie-Landshoff 
flux factor. 
The pomeron, assumed to be scalar, is considered to have
a pointlike coupling to 
a quark pair, which may radiate a gluon, giving rise to $q\bar{q}$
or $q\bar{q}g$ final states, or
to a gluon pair, leading to $q\bar{q}g$ final states through
the fusion of one of the gluons with the photon at the lepton vertex
(Fig.~\ref{fig:feyn-lnmx2}a).
Thus, the two free parameters in the model are the pomeron coupling to
quark ($g_{\pom qq}$) or gluon ($g_{\pom gg}$) pairs.
Two- and three-jet final states are generated using a $y_{min}$
criterion\footnote{For any pair $ij$ of partons, $m_{ij}^2>y_{min}M_X^2$ is 
required to avoid soft and collinear divergences.}
with the value set to 0.03. The value of the QCD scale parameter is chosen
so as to reproduce the values for first order $\alpha_s$ as
measured at PETRA. The Lund
fragmentation scheme \cite{lund} as implemented with its default parameters
in JETSET \cite{jetse} is used.

The RAPGAP and VBLY models are used for correction purposes, since POMPYT
gives a much poorer description of the data, as discussed in the next section.

\section{Results}
 
\subsection{Global shape analysis}

In order to study the dependence of the multihadronic final states on
$M_X$, we calculate in the $\gpom$ rest frame\footnote{Since the outgoing proton is not measured, 
the pomeron direction cannot be unambiguously 
determined. However, its transverse momentum is expected to be small.
We therefore assume the emitted pomeron to be 
collinear with the incident proton, carrying a fraction $x_{\pom}$ of its
momentum.} the
momentum tensor \cite{spher},
$M_{\alpha \beta}=\displaystyle \sum_{i=1}^{n} p_{i\alpha} p_{i\beta}$,
where the sum runs over the $n$ particles (condensates) in a final state
and $\alpha,\beta=X,Y,Z$.     
Diagonalizing the symmetric
tensor $M_{\alpha \beta}$ yields three axes $\vec{n}_k~(k=1,2,3)$ which
give the orientation of the system together with three
eigenvalues 
$\lambda_k= \displaystyle \sum_{i=1}^{n} (\vec{p}_i \cdot \vec{n}_k)^2$,
which can be normalized,
$Q_k= \lambda_k / \displaystyle \sum_{i=1}^{n} |\vec{p}_i|^2$,
such that they
satisfy the relation $Q_1+Q_2+Q_3=1$. By ordering the unit vectors
$\vec{n}_k$ with increasing eigenvalues $Q_1 \leq Q_2 \leq Q_3$, one can define the
variables sphericity $S$, the transverse momenta in $\left<p_{T_{in}}^2\right>$
and out $\left<p_{T_{out}}^2\right>$
of the event plane as follows: 
\[
S=\frac{3}{2} (Q_1+Q_2)=\frac{3}{2}\min_{\vec{n}}\frac{\displaystyle \sum_{i=1}^{n} p_{T_{i}}^{2}}{\displaystyle \sum_{i=1}^{n} p_i^{2}}\;\;\;\;\;\;\;\;(\vec{n} = \vec{n}_3),
\]
\[
\left< p_{T_{in}}^2 \right>=\frac{1}{n}\sum_{i=1}^{n}(\vec{p}_i \cdot \vec{n}_2)^2,
\]
\[
\left< p_{T_{out}}^2 \right>=\frac{1}{n}\sum_{i=1}^{n}(\vec{p}_i \cdot \vec{n}_1)^2,
\]
where $\vec{n_3}$ is the unit vector along the principal axis, the
so called sphericity axis, which minimizes the sum of
the squared transverse momenta. The event plane is defined by
$\vec{n}_2$ and $\vec{n}_3$, while $\vec{n}_1$ defines the direction
 perpendicular to the event plane. Isotropic events are
characterized by $S\sim 1$ and collimated two-jets by $S\sim 0$. In
models with a constant limited transverse momentum, the mean sphericity
values are inversely proportional to the c.m. energy.

Alternatively, one can define the thrust \cite{thrust} axis, denoted
by the unit vector $\vec{n}_T$, as the
direction in space which maximizes the longitudinal momenta, with
 thrust then being defined as
\[
T=\max_{\vec{n}} \frac{\displaystyle \sum_{i=1}^{n} | \vec{p}_i \cdot \vec{n} |} {\displaystyle \sum_{i=1}^{n} |\vec{p}_i |} \;\;\;\;\;\;\;\;(\vec{n} = \vec{n}_T).
\]

Since thrust is a quantity linear in the momenta, it  is
less sensitive to QCD divergences.
Isotropic events are characterized by $T\sim 0.5$ while two-jet events 
tend towards $T\sim 1$.

In Figs.~\ref{fig:angle} and~\ref{fig:sph_thr}, several observed differential 
  distributions (i.e. uncorrected 
for detector and acceptance effects) are presented. Monte Carlo 
studies show that the detector acceptance and resolution 
do not significantly bias the shape of these distributions. 

The observed polar distribution $\frac{dN}{d|cos\theta_S|}$
of the sphericity axis with respect to the
virtual photon direction, for four different
$M_X^{rec}$ intervals, is shown in Fig.~\ref{fig:angle}a. 
In $e^+e^-$ annihilation this distribution is found to be consistent 
with the form
$(1+ cos^2\theta_S)$ which is generally
considered to give experimental evidence of the fermionic nature of
quarks \cite{hanson}. In contrast, the polar distribution of the
sphericity axis measured in the c.m. $\gpom$ system is much
more peaked, and, within the errors, little dependent on
$M_X^{rec}$. This may be interpreted as a consequence of the t-channel quark
propagator in Fig.~\ref{fig:feyn-lnmx2}a for which the Born cross section is
to a good approximation
proportional to $(1-cos^2\theta_S)^{-1}$.

The observed sphericity and thrust 
distributions, in the four $M_X^{rec}$ intervals, are 
shown in Fig.~\ref{fig:sph_thr}. 
As the c.m. energy of the $\gpom$
collision increases, the sphericity (thrust) distribution shrinks towards
 smaller (larger)
values. This is a sign of collimation, i.e. jet formation along the
sphericity or thrust axes. 

\begin{figure}\vspace*{-4cm}
\centerline{\includegraphics[height=24cm]{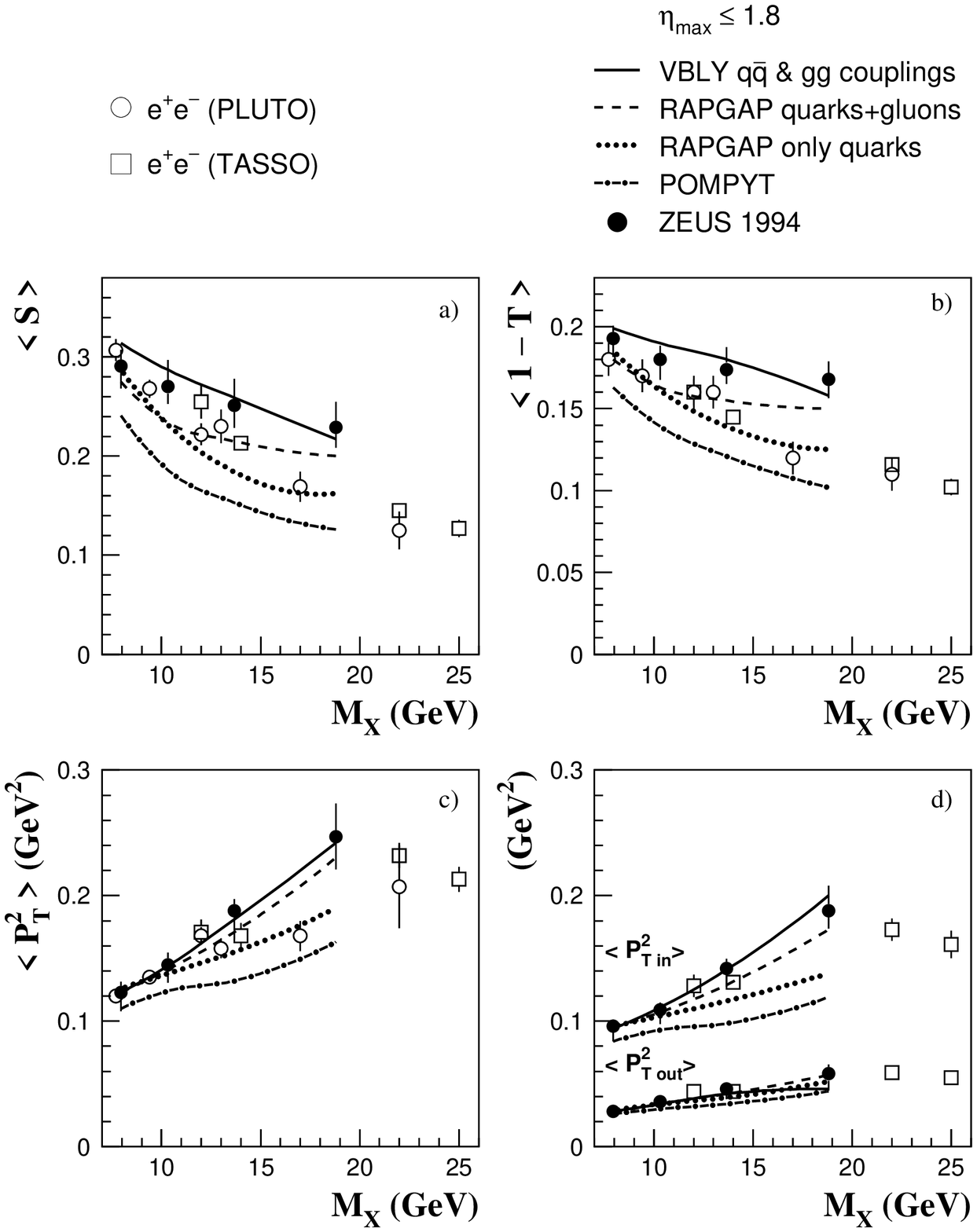}}\vspace{-3cm}
\caption{The mean sphericity, one minus mean thrust, mean squared transverse momenta w.r.t. the sphericity axis and its components in and 
out of the
event plane, plotted as a function of $M_X$. Open dots and squares represent PLUTO and TASSO measurements at DORIS and PETRA, respectively.
Black dots represent the ZEUS data corrected for detector effects to the kinematic
region $5\gev^2 \leq Q^2 \leq 185\gev^2$, $160\gev\leq W \leq 250\gev$ and
$\eta_{max}\leq 1.8$. Note that the ZEUS data and the predictions from the models
are for $\eta_{max}\leq 1.8$.}
\label{fig:zeus_pluto}
\end{figure}

The corrected mean sphericity $\left<S\right>$, 
one minus mean thrust $\left<1-T\right>$ and the 
   mean squared transverse momentum of the final state hadrons w.r.t. the 
  sphericity axis $\left<p_T^2\right>$, as well as its components 
in $\left<p_{T_{in}}^2\right>$ and out $\left<p_{T_{out}}^2\right>$ of
 the event plane, are shown in Fig.~\ref{fig:zeus_pluto} as a function of $M_X$  
    in the kinematic region 
$5 \leq Q^2 \leq 185\gev^2$, 
$160 \leq W \leq 250\gev$  and 
   $\eta_{max} \le 1.8$. Here $\eta_{max}$ is defined as the pseudorapidity of
the most forward generated particle with an energy above $400\mev$.
These results are corrected for all detector
   effects, using Monte Carlo data passed through the detector simulation
  and data reconstruction program.
  The correction factor applied to the measured mean value,
  $\left<X\right>^{meas}_{data}$, is given by   the ratio 
 $\left<X\right>^{gen}_{MC}/\left<X\right>^{rec}_{MC}$,
  where $\left<X\right>^{gen}_{MC}$ and $\left<X\right>^{rec}_{MC}$ 
  are the mean values from the
  generated and reconstructed Monte Carlo data respectively. 
  $\left<X\right>^{gen}_{MC}$ is evaluated at $M_X$ and 
   $\left<X\right>^{rec}_{MC}$ at $M_X^{rec}$.
   The correction factors using the RAPGAP and VBLY event generators
  deviate from unity typically by less than 5\%.
The errors quoted in
this figure include, added in quadrature, systematic uncertainties due to i) model dependences in the
correction procedure, ii) the determination of the boost to the $\gpom$
c.m. system using double angle or scattered lepton variables,
 iii) thresholds for
the definition of condensates, iv) variations in the mass reconstruction procedure
($A_0=0$ and $A_1=1.5$ in previous formula for $M_X^{rec}$)
 and v) variation of the  $\eta_{max}^{rec}$ cut between 1.6 and 2.0. The
systematic error dominates over the statistical.

 Correction to the whole $\eta_{max}$
range is precluded by the fact that the correction becomes model dependent. In fact
while the correction factor within the VBLY model is $\eta_{max}$ independent, in the
RAPGAP model it drops to $0.6$ at the highest mass bin. We interpret this as a consequence
of the different behaviour of the pomeron remnant in the two models. In RAPGAP,
the pomeron remnant, a quark or a gluon, follows 
the proton direction of flight, while in models based on pointlike couplings like VBLY,
the partons at the pomeron vertex are produced at larger angles. 

The mean values $\left<S\right>$ and $\left<1-T\right>$ measured in DIS events
with a large rapidity gap at HERA 
(Figs.~\ref{fig:zeus_pluto}a,~\ref{fig:zeus_pluto}b) decrease with increasing $M_X$.
This decrease is much slower than what would be expected
in a model with constant limited $p_T$ and opposite to isotropic phase space 
predictions \cite{hanson}.
  The mean transverse momentum 
with respect to the sphericity axis (Fig.~\ref{fig:zeus_pluto}c)
increases with increasing $M_X$. This growth is in fact mainly due to the 
transverse momentum component in the event plane, while that out of the
event plane remains fairly constant (Fig.~\ref{fig:zeus_pluto}d). Therefore,
the events become planar. The observed features are similar to those
in $e^+e^-$ annihilation 
as measured by the PLUTO \cite{barreiro}
and TASSO \cite{tassoa} collaborations at DORIS and PETRA 
(see Fig.~\ref{fig:zeus_pluto}) at centre of mass energies comparable to the 
present range of $M_X$.
Overall, the broadening effects measured in DIS final states with a large 
rapidity gap are stronger
than those exhibited by $e^+e^-$ annihilation data. However, we note that the
$\eta_{max}$ requirement precludes a model independent
conclusion valid for the complete diffractive sample.

\subsection{Comparisons with models}
 
In the previous section we have presented evidence
that the DIS large rapidity gap multihadronic final states become planar
at high masses. POMPYT, for which 
broadening effects are given by the leading log approximation for 
gluon bremsstrahlung from the quark lines,
 does not describe the data.
Not only does it fail to reproduce the $M_X^{rec}$ distribution
(not shown), but more important, the expected mean values for the event shape
variables studied 
are much lower than the data (see Fig.~\ref{fig:zeus_pluto}).

\begin{figure}\vspace{-4cm}
\centerline{\includegraphics[height=27cm]{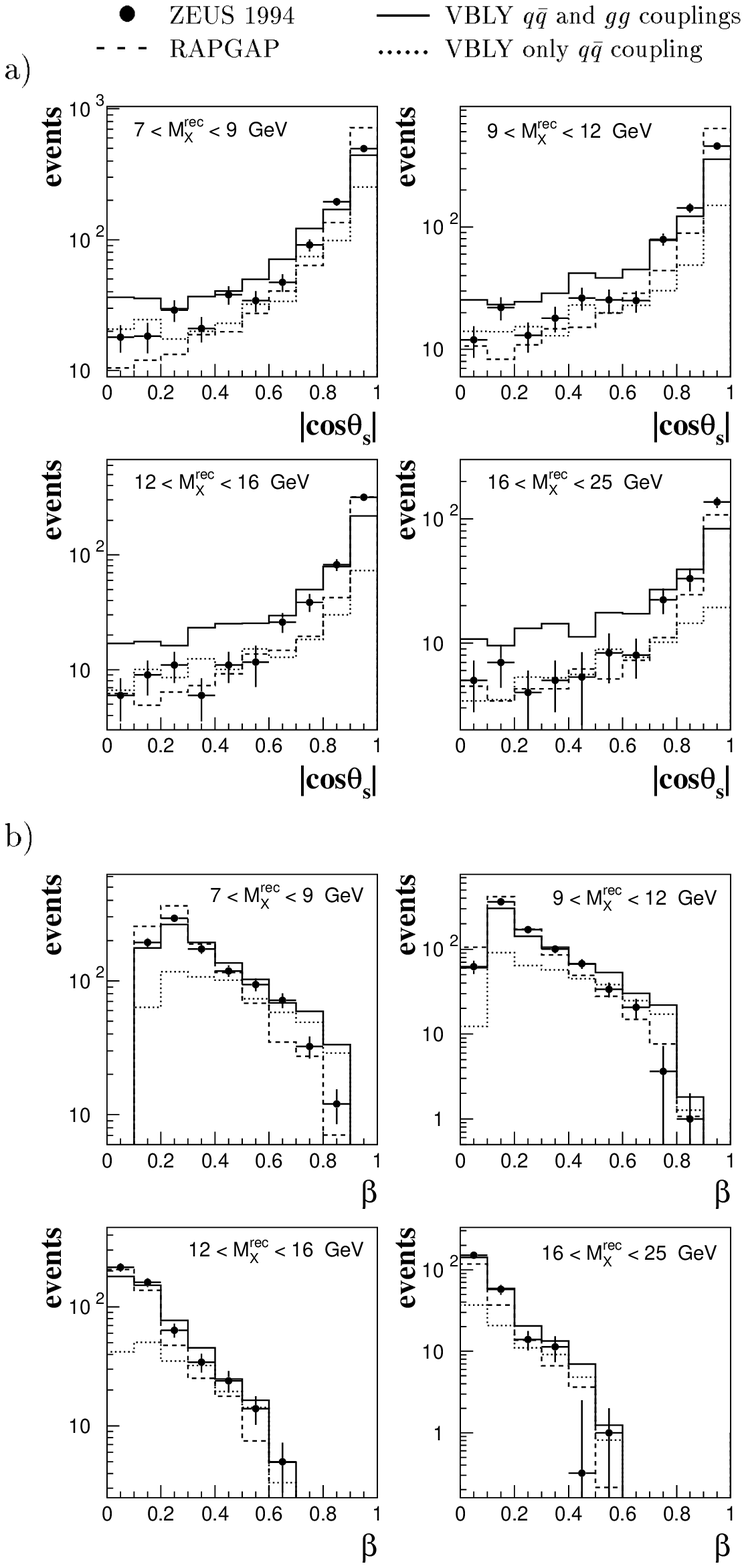}}\vspace{-2.5cm}
\caption{Measured polar distribution of the sphericity axis with respect to the
virtual photon direction in the $\gpom$ rest frame (a), and measured $\beta$ 
distribution (b). The results of the
VBLY model are shown as solid lines, with the quark contribution from this model
shown as dotted lines. RAPGAP expectations are represented by the dashed lines.
The global normalization of the predictions is adjusted so as to
match the measured number of events in the whole mass range $7 < M_X^{rec}<25\gev$.}
\label{fig:angle}
\end{figure}

In the VBLY approach \cite{vermaseren}, the two free parameters, 
$g_{\pom qq}$ and $g_{\pom gg}$, are directly related to the 
fraction of final state events coming from the gluon coupling 
as well as to the 
total diffractive cross section. This fraction was
determined by fitting to a set of observed distributions such as sphericity,
thrust, $\left<p_{T_{in}}^2\right>$, $\left<p_{T_{out}}^2\right>$ 
and $|\cos\theta_S|$ as well as the 
 $\beta$ distributions shown in Fig.~\ref{fig:angle}b,
 all of them plotted in the
four mass intervals considered before. 
 The parameters in RAPGAP have been 
tuned to describe the H1 diffractive structure 
function $F_2^D$ in~\cite{f2dh1} and no additional tuning has been 
attempted in this analysis.   
  
\begin{figure}\vspace{-4cm}
\centerline{\includegraphics[height=27cm]{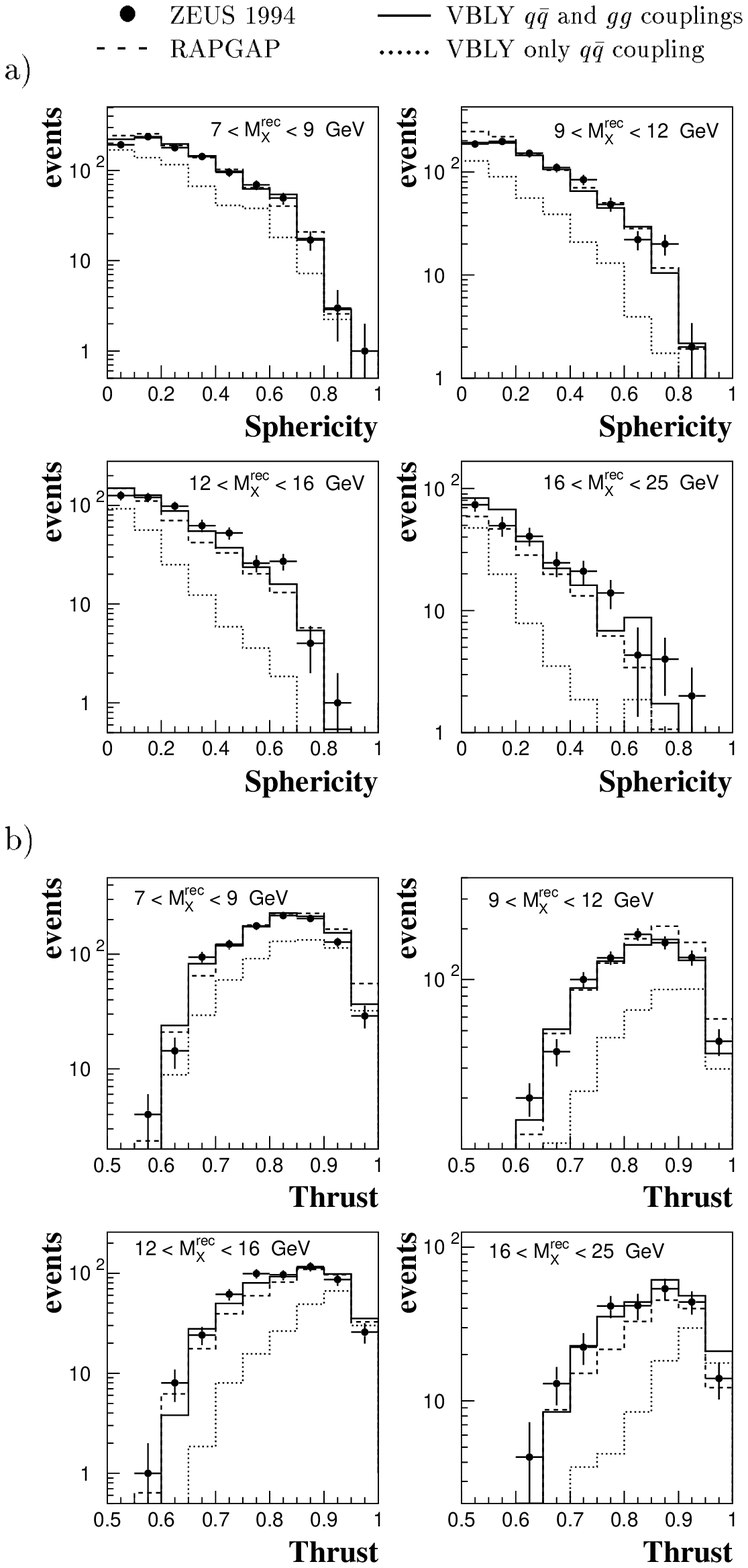}}\vspace{-2.5cm}
\caption{Measured sphericity (a), and thrust distributions (b) for large rapidity gap DIS events in four
$M_X^{rec} $ intervals. The results of the VBLY model are shown as solid lines, with the quark contribution from this model shown as dotted lines. 
RAPGAP expectations
are represented by the dashed lines. 
The global normalization of the predictions is adjusted so as to
match the measured number of events in the whole mass range $7 < M_X^{rec}<25\gev$.}
\label{fig:sph_thr}
\end{figure}

The results of the VBLY fits (solid lines) and the expectations from 
RAPGAP (dashed lines) are shown in 
Figs.~\ref{fig:angle},~\ref{fig:sph_thr}, and~\ref{fig:zeus_pluto}. 
The global normalization in Figs.~\ref{fig:angle} and~\ref{fig:sph_thr}
is determined by the total number of events in the
whole mass range $7 < M_X^{rec}<25\gev$.
Both models 
give a reasonable description of the data. 
Some discrepancies are observed in the
$|\cos\theta_S|$ distribution in VBLY. Although in the VBLY model the pomeron
coupling to quarks and gluons is pointlike, the $\beta$ distributions,
presented in Fig.~\ref{fig:angle}b, are also reasonably described.
The observed shape at small $\beta$ is generated through the coupling 
of the pomeron to gluon pairs.

RAPGAP with only a quark density in the pomeron (dotted line in
Fig.~\ref{fig:zeus_pluto}) fails to describe the data.
Similarly, VBLY with only the pomeron to quark pair coupling  
cannot describe the shapes of
the sphericity, thrust and $\beta$ distributions (dotted line in
Fig.~\ref{fig:angle} and~\ref{fig:sph_thr}). 
Note that the POMPYT prediction shown in Fig.~\ref{fig:zeus_pluto} is similar 
to that of VBLY with only the quark pair coupling (not shown
in Fig.~\ref{fig:zeus_pluto}). 

The implementation of the  
hard gluon bremsstrahlung process in the MC models 
is insufficient to account for the broadening effects observed in the data.
Therefore, the need to include either the pomeron to gluon pairs
coupling in VBLY or the gluon density in RAPGAP is very clear. 
The exact fraction of gluon-induced events is model dependent and it is 
sensitive to the $y_{min}$ and $p_T$ cuts imposed on the LO matrix elements
used. However, in both models a significant fraction of gluon-induced events
is required to describe the data. In the kinematic range under study, this
fraction is 50\% for VBLY and 30\% for RAPGAP.

\section{Conclusions}

We have studied global event shapes in large rapidity gap NC DIS events at HERA 
in the kinematic
range $5 \leq Q^2 \leq 185\gev^2$, $160 \leq W \leq 250\gev$
 and $\eta_{max}\leq 1.8$. These events are generally interpreted as due to
pomeron exchange. We have
investigated the dependence of the event shape variables  sphericity, thrust
and transverse momenta squared in and out of the event plane with
$M_X$, the c.m. energy of the $\gpom$ collision. The polar distribution
of the sphericity axis is very peaked and little dependent on $M_X$.
We find that with increasing $M_X$ the large rapidity gap events become more collimated and
planar. 
Broadening effects in the DIS final states with a large rapidity gap call for a
mechanism in addition to hard gluon bremsstrahlung. This can be achieved
in models where the pomeron has a partonic structure by including 
a gluon density in the pomeron, as in RAPGAP, or through a 
pointlike coupling of 
the pomeron to quark and gluon pairs, as in VBLY. 
A significant gluon component of the pomeron is necessary for the models to
describe the data.
 
\vspace{2cm}
\noindent {\Large\bf Acknowledgments}

    The strong support and encouragement by the DESY Directorate have 
    been invaluable. The experiment was made possible by the
    inventiveness and diligent efforts of the HERA machine group. 
    We acknowledge the support of the DESY computing and network services.

    The design, construction and installation of the ZEUS detector have
    been made possible by the ingenuity and dedicated efforts of many 
    people from the home institutes who are not listed here. Their 
    contributions are acknowledged with great appreciation. 

    We would like to thank J. Vermaseren and F.J. Yndur\'ain
    for very valuable discussions.


\clearpage

\clearpage

\clearpage

\clearpage

\end{document}